\documentclass[bibyear]{aa}  

\usepackage{graphicx}
\usepackage[version=4]{mhchem}
\usepackage{color,soul} 

\usepackage{tabularx}
\usepackage{multirow}
\usepackage{booktabs}
\usepackage{siunitx}
\usepackage{natbib}

\usepackage{txfonts}
\usepackage[pdftex,plainpages=false,
    pdfpagelabels,
    pdfusetitle,  
    bookmarksnumbered,
    linkbordercolor={1 1 1},
    hidelinks=true,
    colorlinks={ true },
]{hyperref}

\newcommand{\eg}{e.\,g.}

\newcommand{\kL}{\ensuremath{k_{\text{L}}}}

\usepackage{xcolor}
\definecolor{DVIgreen}{HTML}{00A64F}

\usepackage{stfloats}

\begin{document}

    \title{Measurements and simulations of rate coefficients for the deuterated forms of the \ce{H2+ + H2} and \ce{H3+ + H2} reactive systems at low temperature}

   \titlerunning{Rate coefficients for deuterated forms of \ce{H2+ + H2} and \ce{H3+ + H2}}


   \author{Miguel Jim{\'e}nez-Redondo \and
          Olli Sipil\"{a} \and
          Pavol Jusko \and
          Paola Caselli
          }

   \institute{Max Planck Institute for Extraterrestrial Physics, Gie{\ss}enbachstra{\ss}e 1, 85748 Garching, Germany\\ \email{pjusko@mpe.mpg.de}}

   \date{}

  \abstract
   {The rate coefficients
   of various isotopic variations of the \ce{H2+ + H2} and \ce{H3+ + H2} reactions in the 10--250~K temperature range were 
   measured using a cryogenic 22 pole radio frequency ion trap. 
   The processes involving diatomic ions were found to behave close to the Langevin rate, whereas temperature-dependent 
   rate coefficients were obtained for the four isotopic exchange processes with triatomic ions. Fitting the 
   experimental data using a chemical code allowed us in specific cases to constrain rate coefficients that were not directly measured 
   in the ion trap. The reported rate coefficients suggest a more efficient hydrogenation of deuterated \ce{H3+} forms than
   usually assumed in astrochemical models, which might affect deuteration rates in warmer environments.}

   \keywords{Astrochemistry, Molecular processes, Methods: laboratory: molecular, Methods: numerical, Astronomical data bases,
   ISM: molecules %
               }

   \maketitle
%
\section{Introduction}
\ce{H3+} is the most important molecular ion in astrochemistry \citep{Oka2013}. It is formed by collisions of \ce{H2+} with molecular hydrogen,
\begin{equation}
    \ce{H2+ + H2 -> H3+ + H}.
\end{equation}
In dense and cold molecular clouds, \ce{H3+} serves as the universal protonator. It reacts with an atom or molecule M,
\begin{equation}
    \ce{H3+ + M -> MH+ + H2},
\end{equation}
and initiates
a chain of ion-neutral reactions that causes the formation of most
interstellar molecules \citep{McCall2000,Tennyson1995}.

At the low temperatures ($\sim 10$~K) of molecular clouds, the zero point energies (ZPE) of the reactants and products 
can play an important role in favoring particular chemical pathways. 
Because of this, deuterium is more efficiently incorporated into molecules,
where it produces enhancements several times higher than the galactic D/H ratio \citep{Millar1989,Millar2002}. 
For \ce{H3+}, isotopic exchange reactions with HD can lead to the deuteration of the triatomic ion,
    \begin{equation}
    \ce{H3+ + HD <--> H2D+ + H2}
    \end{equation}
    \begin{equation}
    \ce{H2D+ + HD <--> D2H+ + H2}
    \end{equation}
    \begin{equation}
    \ce{D2H+ + HD <--> D3+ + H2}.
    \end{equation}
The low exothermicity of these processes ($\sim 200$~K) means that the reverse reactions may still be relevant even under 
low-temperature conditions. Understanding the isotopic effects on the chemistry of \ce{H2+} and \ce{H3+} is therefore key 
to modeling the deuterium fractionation in molecular clouds \citep[e.g.][]{Caselli2003,Walmsley2004,Caselli2019}, 
protoplanetary disks \citep[e.g.][]{Willacy2009}, and other environments, including comets
\citep[e.g.][]{Drozdovskaya2021} and external galaxies \citep[e.g.][]{Bayet2010,Shimonishi2021}.

Isotopic effects are not only relevant in astrochemistry. The balance of isotopic exchange processes is key for the ionic 
chemistry of \ce{H2}/\ce{D2} cold plasmas \citep{Redondo2011}, and enhanced rates for deuteration reactions can have a 
significant role in the efficiency of fusion plasmas \citep{Verhaegh2021}. 
Substantial effort from the experimental and theoretical point of view, has therefore been devoted to the characterization of the \ce{H2+ + H2} and \ce{H3+ + H2} isotopic systems.

The exoergic reaction of \ce{H2+} with \ce{H2} to form \ce{H3+} has been extensively studied mainly through the use of merged-beam techniques.
The cross section of this process has been characterized for energies ranging from a few meV \citep{Glenewinkel1997,Allmendinger2016a} 
to a a few eV \citep{Pollard1991,Savic2020} and was found to be well described by the Langevin model for energies below 1~eV. 
Similar studies were also performed for isotopically substituted reactants, such as the work of \cite{Krenos1976} 
on \ce{H2+ + D2} and \ce{D2+ + H2} and \cite{Douglass1977} on \ce{HD+ + D2} and \ce{D2+ + HD}. More recently, 
the behavior of the cross section at very low temperatures (< 1~K) was examined for \ce{H2+ + H2} 
\citep{Allmendinger2016b}, \ce{H2+ + HD} \citep{Hoveler2021a}, and \ce{H2+ + D2} \citep{Hoveler2021b}. 
The enhancement of the cross section compared to the Langevin model found in these conditions \citep{Allmendinger2016b,Hoveler2021b} 
matches the theoretical prediction of \cite{Dashevskaya2016}. In a recent study, \cite{Merkt2022} provided branching 
ratios for all isotopic variations of the \ce{H2+ + H2} system for this temperature range. A variety of theoretical 
efforts employing different approaches to model this process can be found in the literature, often using quasiclassical 
trajectory methods \citep{Krenos1976,Eaker1985,Sanz-Sanz2015}, although the system was also studied with 
a quantum mechanical treatment \citep{Baer1990}. 

The first experimental low-temperature studies of the isotopic exchange variations in \ce{H3+ + H2} were 
performed in selected ion flow tube experiments \citep{Adams1981,Giles1992}. The corresponding rate 
coefficients and branching ratios were determined at temperatures of 80 and 300~K. More recent low-temperature studies, such as the deuteration of \ce{H_n+} clusters, including \ce{H3+} \citep{Paul1996}, 
have employed ion trap setups. \cite{Gerlich2002} studied the temperature-dependent reaction rate 
of \ce{D3+} with \ce{H2} down to 10~K in a 22 pole trap. The same setup was used to characterize 
the \ce{H3+ + HD} reaction \citep{Gerlich2002a}, and the reactions of the deuterated ions subsequently 
produced inside the trap with HD using a kinetic model fit to the experimental data. 
The controversial set of rate coefficients that was obtained was lower than expected, and in a follow-up 
study, \cite{Hugo2009} repeated this experiment and obtained values that were about four times higher than in 
the former work. The experimental results of \cite{Hugo2009} were complemented by a spin-state resolved
microcanonical model that agreed well with their measured rates. None of these 
experiments resolved the spin state of the reacting ion because it is experimentally difficult to prepare 
ions in defined states inside an ion trap, with the exception of the \ce{H+ + H2} reaction, where \ce{H+} 
lacks electronic spin and \ce{p-H2} is easily produced \citep{Gerlich2013}. From the theoretical 
point of view, a number of studies have focused on the possible hopping and exchange processes 
and their possible constraints. Potential energy surfaces of \ce{H5+} were calculated by 
some groups \citep{Xie2005,Aguado2010}. The role of spin states was highlighted by \cite{Gerlich2006} and \cite{Gomez-Carrasco2012}, 
while the balance between proton hop and 
exchange was recently examined using ring polymer molecular dynamics \citep{Suleimanov2018}.

Despite this extensive research, experimental data on the rate coefficients of some of these 
processes at the temperatures relevant for astrochemistry are still lacking. In this work, we present measurements of the rate coefficients 
of several of the isotopic variations of the \ce{H2+ + H2} and \ce{H3+ + H2} systems in the range of 10--250~K using 
a cyogenically cooled 22 pole ion trap setup. The experiments are complemented by simulations with a chemical code 
in order to fit the measured data, and we attempted to derive additional rate coefficients, which were not directly 
accessible experimentally.

\section{Experiment}
\subsection{Experimental setup}

The experimental setup consisted of a 22 pole radio frequency (rf) trap with stainless-steel rods (5~cm long and 1~mm in diameter) 
placed around a circle with a 5~mm inscribed radius, and surrounded by ring electrodes. The setup was described in detail in
\cite{Jusko2023}, and we only provide a brief description here. 
The trap was mounted on top of the second stage of a helium cryostat, allowing the cooling of the setup down to 4~K.
Ions were generated in a Gerlich-type storage ion source (SIS) by electron bombardment \citep{Gerlich1992}, using \ce{H2},
\ce{D2}, or a mixture of both as precursor gases. The ions were subsequently
guided through a first quadrupole acting as a mass filter, which allowed us to select the ions according to their 
mass-to-charge ratio ($m/z$). The ion beam was deflected with a quadrupole bender before it arrives in the 22 pole trap, 
where an intense pulse of He buffer gas was used to slow it down and relax the internal excitation. This
allowed us to trap the ions and cool them down to the temperature of the trap. 
After an established trapping time, the ions were extracted from the trap through the output electrode
and were subsequently detected (counted) with a quadrupole mass spectrometry system.

The reactant gas (normal \ce{H2} or \ce{D2}) was continuously injected into
the trap in order to induce the desired ion-molecule reaction. The time evolution of the 
different ionic species was then recorded by performing several experiments
in which the trap was emptied after various storage times.
The rate coefficients of interest can be readily derived from the exponential decay of the primary
ion signal, but a more robust result can be obtained by repeating the experiment for 
several different values of the number density of the neutral 
(usually around $10^{10}$~cm$^{-3}$) and performing a linear fit of the decay 
rate against number density, as described in \cite{Jusko2023}.

\subsection{Measured rate coefficients}

The rate coefficients of the reaction of most isotopologs of \ce{H2+} and \ce{H3+} with \ce{H2 / D2} were determined 
using the above-mentioned method. However, issues with mass selectivity prevented us from studying some of these processes. 
Since ions are discriminated according to their mass-to-charge ($m/z$) ratio, this prevented us from unequivocally selecting 
the {\ce{H2D+}} ion in the first quadrupole,
since its $m/z$ of 4 is also shared by {\ce{D2+}}, 
which will be produced in the ion source in significant amounts for any precursor mixture containing {\ce{D2}}. 
This conflict in masses also prevented us from studying the {\ce{D2+ + H2}} process, as the {\ce{H2D+}} 
product branch would contribute to the monitored 4~$m/z$ signal and prevent the determination of the decay rate for \ce{D2+}.

When the primary ions were diatomic (\ce{H2+}, \ce{D2+} or \ce{HD+}), 
the collisions of the energetic (up to eV) ion beam 
with the He buffer gas during the trapping process
correspondingly resulted in the production of \ce{HeH+} and \ce{HeD+}. This was not observed for 
the triatomic ions \ce{H3+} and \ce{D2H+}, and even though the conflicting masses of \ce{D3+} and \ce{HeD+} prevented a 
straightforward dismissal, the behavior of the system indicated that this is also the case for \ce{D3+}. 
The reaction of the diatomic ions with He is quite endothermic for ground-state reactants \citep[by 0.83~eV in the case 
of \ce{H2+},][]{Glosik1994}, and it is consequently only observed to occur during the trapping process and not 
during the subsequent storage time. Since these \ce{HeX+} ions were co-trapped with the corresponding primary ion, 
we were able to simultaneously determine both rate coefficients in reaction with the neutral of interest.

\begin{figure}
\includegraphics[width=\columnwidth]{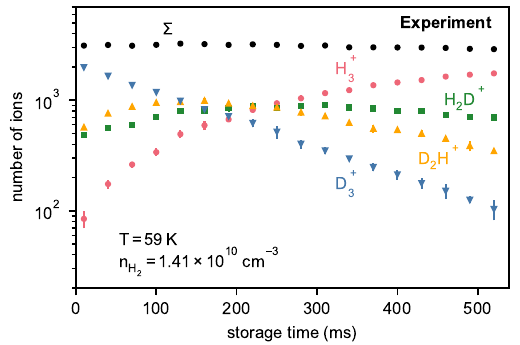}
\caption{Experimental number of ions in the trap as a function of trapping time for
  the \ce{D3+ + H2} experiment at 27 K. $\Sigma$ represents the sum of all ions. \ce{H2D+} and \ce{D2H+} undergo subsequent reactions 
  with \ce{H2}, eventually leading to the formation of \ce{H3+}.}
\label{fig_exp_time}
\end{figure}

\begin{figure}[ht]
\resizebox{\hsize}{!}{\includegraphics{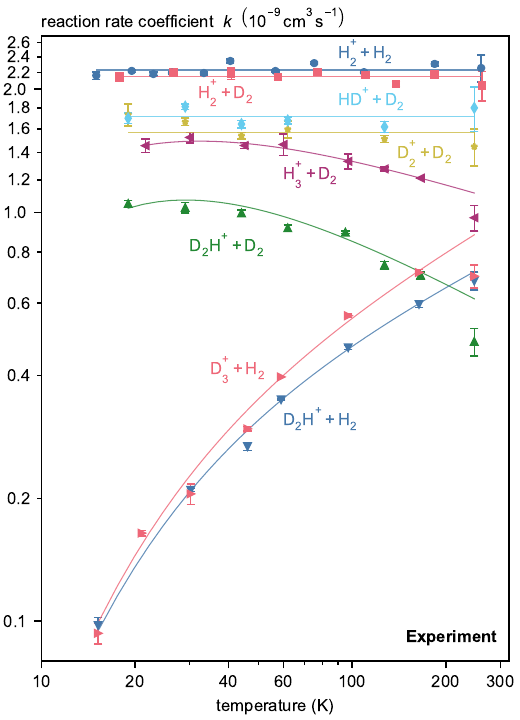}}
\caption{Rate coefficients as a function of temperature for the eight reactions studied in the 22 pole trap experiments. 
  The symbols correspond to the measurements, and the lines represent the fit to the Arrhenius-Kooij formula with the parameters given 
  in Table\,\ref{tab:kida}.}
\label{fig_exp}
\end{figure}

For many of the studied ion-molecule pairs, the product ions underwent subsequent reactions with the neutral present 
in the trap. This is illustrated in Fig.\,\ref{fig_exp_time} for the case of \ce{D3+ + H2}. In this particular case, 
the reaction proceeded until full hydrogenation was achieved. The experimental rate coefficients (see Fig.\,~\ref{fig_exp}) 
were only calculated from the decay of the primary ion when it reacted with the neutral, but the full data of the experiment 
can be processed with the help of a kinetic model (see below). Due to the exothermicity of some of the ion-molecule processes we studied, 
the products of these reactions can possess some degree of internal excitation, unlike the buffer-cooled primary ions. 
This possibility must be taken into account when processing the full data of these experiments.

The experiments presented here did not resolve the spin state of the ions. Nevertheless, spin
relaxation might take place in the trap for the less efficient reactions, which are those
involving triatomic ions. However, according to \cite{Hugo2009}, collisions modifying the ionic nuclear
spin are forbidden for all studied reactions except \ce{D2H+ + D2}, where they are
significantly less efficient than the reactive pathway. 
This means that for all experiments, the spin-state distribution of the ions
will be the one resulting from their production in the ion source, which is not characterized.

The experimentally determined rate coefficients are shown in Fig.\,\ref{fig_exp}. The error bars 
represent the standard deviation of the data fit. Measurements at the lowest and 
highest temperature usually present an increased uncertainty associated with the freezing of 
the reactant gas in the colder parts of the trap, which affects the reliability of the pressure 
measurements, or the presence of residual species such as water giving place to parasitic 
reactions, respectively. Overall, the main source of error for these measurements is the 
uncertainty in the derivation of the number 
density, which is estimated to be $\sim 20\%$ \citep{Jusko2023}. 

A clear separation was observed between the efficient reactions 
involving diatomic ions, with exothermicities around 1~eV, 
and most of the isotopic exchange processes of the triatomic ions, which 
are slower by a factor of 2--4 times.
The reactions of the diatomic ions do not show a significant temperature dependence, including those 
of \ce{HeH+} and \ce{HeD+} with \ce{H2}, which were omitted from the plot for clarity (see Appendix~\ref{appendix:HeX}). 
Only in the case of the
\ce{D2+ + D2} reaction is it possible to observe a decay in the rate coefficient of about 10\% between 20 and 250 K,
but the values are still within the expected dispersion for these measurements.
In contrast, the effect of temperature is clearly noticeable for all reactions involving triatomic ions. 
The starkest temperature dependence is observed for the hydrogenation reactions, which, depending 
on the product channel, are endothermic by 234/342~K in the case of \ce{D3+ + H2} and 187/340~K 
for \ce{D2H+ + H2}, with ZPE taken from \cite{Ramanlal2003}. The rate coefficient for these
two processes grows by almost an order of magnitude between 15 and 250~K. The fact that these reactions are still observed
to proceed at a moderate rate even at the lowest temperatures we studied is likely due to the internal energy of ortho-\ce{H2},
175~K, which helps offset the endothermicity of the process.
The deuteration reactions \ce{H3+ + D2} and \ce{D2H+ + D2}, which are in turn exothermic by 153/340~K and 155~K, respectively,
were found to be significantly more efficient at 
low temperatures, with a somewhat stable rate coefficient at low temperatures and a noticeable decrease above $\sim 60$~K, 
leading to a reduction in the rate coefficient of $\sim 30\%$ at 250~K compared to 20~K.

The temperature dependence of these rate coefficients was fit according to the Arrhenius-Kooij formula, 
so that we were able to directly use it with tools based on the KIDA data format \citep{Wakelam2012}.
The fit is represented by the solid lines in Fig.\,~\ref{fig_exp}. The parameters of the fit, together with the uncertainties derived 
from the statistical analysis, are collected in Table~\ref{tab:kida}.

\begin{table}[ht]
    \caption[]{Parameterized experimentally determined reaction rate coefficients from Fig.\,\ref{fig_exp}
    and Fig.\,\ref{fig_exp_He_only} between 15--280 K (reactions with \ce{H2}) and 20--280 K (reactions with \ce{D2}).
    }\label{tab:kida}
    \begin{center}
    \sisetup{separate-uncertainty}
    \begin{tabular}{lrrr} 
       \toprule
         Reaction  & \multicolumn{1}{c}{$A^a$} & \multicolumn{1}{c}{$B$} & \multicolumn{1}{c}{$C^c$}  \\
       \midrule
\ce{H2+ + H2}   & $22.30 \pm 0.16$  &                   &                   \\
\ce{H2+ + D2}   & $21.54 \pm 0.16$  &                   &                   \\
\ce{HD+ + D2}   & $17.16 \pm 0.34$  &                   &                   \\
\ce{D2+ + D2}   & $15.65 \pm 0.29$  &                   &                   \\
\ce{D2H+ + H2}  & $8.21 \pm 0.40$   & $0.36 \pm 0.07$   & $16.71 \pm 3.56$  \\
\ce{D3+ + H2}   & $10.22 \pm 1.03$  & $0.41 \pm 0.12$   & $16.92 \pm 5.96$  \\
\ce{H3+ + D2}   & $10.98 \pm 0.13$  & $-0.25 \pm 0.04$  & $8.06 \pm 2.80$   \\
\ce{D2H+ + D2}  & $5.94 \pm 0.52$   & $-0.44 \pm 0.13$  & $12.86 \pm 5.16$  \\
\ce{HeH+ + H2}  & $17.45 \pm 0.23$  &                   &                   \\
\ce{HeD+ + H2}  & $15.74 \pm 0.23$  &                   &                   \\

      \bottomrule
    \end{tabular}
    \end{center}
    \tablefoot{$a$ -- in $10^{-10}\;\text{cm}^3\,\text{s}^{-1}$; $c$ -- in $\text{K}$.
    Fit in the form of the Arrhenius-Kooij formula \citep{Wakelam2012} $k(T) = A(T/300)^B \exp(-C/T)$. 
    The uncertainties correspond to the statistical errors of the least-squares fit.}
\end{table}

A comparison of the experimental rate coefficients with some of the previously
available values is shown in Table~\ref{tab:k_comp}. 
Overall, the measurements presented here match
previous works such as the selected-ion flow-tube (SIFT) experiments of \citet{Giles1992} at 80 and 
300~K. Only in very specific cases are values for the thermal rate
coefficients at cold (10--20~K) temperatures available in the literature. 
For \ce{H2+ + H2}, merged-beam experiments such as those by
\cite{Allmendinger2016a} and \cite{Savic2020} showed that the cross section
for this process at low energies is very close to the Langevin value.
The low rate coefficient obtained for \ce{D3+ + H2} is in line with
the results of another merged-beam experiment by \cite{Gerlich2002a}.
Finally, the value for the \ce{H3+ + D2} rate coefficient obtained here
is close to the value measured by \cite{Paul1996} in a 22 pole trap,
which presented significant uncertainty.

\begin{table*}[ht]
    \caption[]{Comparison of the rate coefficients obtained in this work with values available in the 
    literature below 100~K. All values are in $10^{-10}\;\text{cm}^3\,\text{s}^{-1}$.
    }\label{tab:k_comp}
    \begin{center}
    \sisetup{separate-uncertainty}
    \begin{tabular}{l S[table-format = 2.1] S[table-format = 4.1] S[table-format = 4.1] S[table-format = 4.1] S[table-format = 4.1]}
    
       \toprule
         Reaction      &  {\kL} & {this work (20~K)}      &      {other works ($\sim 20\;\text{K}$)} & {this work (80~K)} & {\cite{Giles1992} (80~K)} \\
       \midrule
\ce{H2+ + H2}   & 21.0       & 22.3       & 21.0 $\;^{a,b,c}$     & 22.3       &             \\
\ce{H2+ + D2}   & 18.0       & 21.5       & 18.0 $\;^{c}$         & 21.5       &             \\
\ce{HD+ + D2}   & 15.9       & 17.2       &                       & 17.2       &             \\
\ce{D2+ + D2}   & 14.7       & 15.7       &                       & 15.7       &             \\
\ce{D2H+ + H2}  & 17.6       & 1.4        &                       & 4.2        & 6.0         \\
\ce{D3+ + H2}   & 17.2       & 1.4        & 1.2 $^{d}$            & 4.8        & 5.2         \\
\ce{H3+ + D2}   & 15.9       & 14.4       & 17.0 $^{e}$           & 13.8       & 14.0        \\
\ce{D2H+ + D2}  & 14.0       & 10.4       &                       & 9.1        & 8.7         \\

      \bottomrule
    \end{tabular}
    \end{center}
    \tablefoot{$a$ -- \cite{Allmendinger2016a}, $b$ -- \cite{Savic2020},
    $c$ -- \cite{Hoveler2021b},
    $d$ -- 2 meV collision energy  \citep{Gerlich2002}, 
    $e$ -- at 10~K \citep{Paul1996}. 
    The values displayed for this work correspond to the Arrhenius-Kooij fit 
    (see Table~\ref{tab:kida}) at the temperature of interest. Langevin reaction rates \kL\ are given for 
    comparison \citep[polarizabilities from][]{Milenko1972}.
    }
\end{table*}

\section{Simulations of the \ce{H3+ + H2} reacting system}

We carried out simulations to corroborate the rate coefficient values pertaining to the $\rm H_3^+ + H_2$ system 
(including combinations of all deuterated forms) measured in the laboratory, and to investigate whether we can obtain 
predictions for rate coefficients that cannot be measured directly in the experiments. To this end, we employed the 
gas-grain chemical code pyRate \citep[presented in detail in, e.g.,][]{Sipila15a}, although in the present case, we only 
required a simple gas-phase description of the chemistry. In what follows, we give a detailed account of the simulations 
performed for the present paper, including a discussion of the choice of initial conditions, and setting of the simulation parameters 
to correspond to the experimental setup.

\subsection{Initial conditions}

Our approach to the chemical simulations was to define an initial set of rate coefficients for the reaction system presented in 
Table\,\ref{tab:kida}, representing a first guess, and to vary these rate coefficients until a reasonable agreement with 
the experimental results was met (see Sect.\,\ref{ss:exampleSystem}). 
The fiducial rate coefficient values were obtained based on the values given in Table\,B.1 of \citet{Sipila17b}. 
There, new fits of rate coefficients for the \ce{H3+ + H2} reacting system, based on the data of \cite{Hugo2009}, were presented. 
These reactions explicitly involve the spin states of the participating species. Our experiments did not 
distinguish the spin states, however, and therefore, they were removed from the simulation network. 
We produced a simplified network by summing the rate coefficients of each reaction over the possible product channels, and we then averaged over the reactant channels assuming that the possible spin forms of a given species are distributed according to their high-temperature 
statistical weights. Table\,\ref{tab:averagedRateCoeffs} shows the resulting set of rate coefficients at four selected temperatures.

\begin{table*}
\caption{Averaged rate coefficients, based on the spin-state resolved data given in \citet{Sipila17b}, for four selected temperatures.}             
\label{tab:averagedRateCoeffs}      
\centering          
\begin{tabular}{c l c l c c c c}
\toprule      
Number & \multicolumn{3}{c}{Reaction} & $k\;(21.6\,\rm K)$ & $k\;(30.0\,\rm K)$ & $k\;(60.0\,\rm K)$  & $k\;(97.0\,\rm K)$\\
\midrule
   1 & \ce{H3+ + HD } & $\to$ & \ce{ H2D+ + H2}  & 7.56(-10) & 7.58(-10) & 7.56(-10) & 7.49(-10) \\
   2 & \ce{H3+ + D2 } & $\to$ & \ce{ H2D+ + HD}  & 6.41(-10) & 6.71(-10) & 7.52(-10) & 8.21(-10) \\
   3 & \ce{H3+ + D2 } & $\to$ & \ce{ D2H+ + H2}  & 9.18(-10) & 8.88(-10) & 8.20(-10) & 7.73(-10) \\
   4 & \ce{H2D+ + H2} & $\to$ & \ce{ H3+ + HD}   & 9.69(-11) & 1.04(-10) & 1.22(-10) & 1.38(-10) \\  
   5 & \ce{H2D+ + HD} & $\to$ & \ce{ H3+ + D2}   & 1.17(-13) & 3.32(-13) & 1.41(-12) & 2.60(-12) \\  
   6 & \ce{H2D+ + HD} & $\to$ & \ce{ D2H+ + H2}  & 5.64(-10) & 5.54(-10) & 5.33(-10) & 5.17(-10) \\  
   7 & \ce{H2D+ + D2} & $\to$ & \ce{ D2H+ + HD}  & 1.14(-09) & 1.14(-09) & 1.16(-09) & 1.17(-09) \\
   8 & \ce{H2D+ + D2} & $\to$ & \ce{ D3+ + H2}   & 2.38(-10) & 2.26(-10) & 2.00(-10) & 1.84(-10) \\
   9 & \ce{D2H+ + H2} & $\to$ & \ce{ H3+ + D2}   & 2.98(-15) & 3.22(-14) & 7.41(-13) & 2.50(-12) \\  
  10 & \ce{D2H+ + H2} & $\to$ & \ce{ H2D+ + HD}  & 5.29(-11) & 8.12(-11) & 1.97(-10) & 3.25(-10) \\  
  11 & \ce{D2H+ + HD} & $\to$ & \ce{ H2D+ + D2}  & 2.04(-13) & 6.16(-13) & 4.47(-12) & 1.27(-11) \\  
  12 & \ce{D2H+ + HD} & $\to$ & \ce{ D3+ + H2}   & 3.26(-10) & 3.10(-10) & 2.72(-10) & 2.46(-10) \\
  13 & \ce{D2H+ + D2} & $\to$ & \ce{ D3+ + HD}   & 1.08(-09) & 1.07(-09) & 1.04(-09) & 1.02(-09) \\
  14 & \ce{D3+ + H2 } & $\to$ & \ce{ H2D+ + D2}  & 2.76(-15) & 4.78(-14) & 2.56(-12) & 1.38(-11) \\  
  15 & \ce{D3+ + H2 } & $\to$ & \ce{ D2H+ + HD}  & 1.96(-10) & 2.84(-10) & 5.22(-10) & 7.05(-10) \\
  16 & \ce{D3+ + HD } & $\to$ & \ce{ D2H+ + D2}  & 3.54(-13) & 1.72(-12) & 1.78(-11) & 5.01(-11) \\

\bottomrule            
\end{tabular}
\tablefoot{Rate coefficients $k$ in $\text{cm}^3\,\text{s}^{-1}$. The notation $a(-b)$ means $a \times 10^{-b}$.}  
\end{table*}

\begin{figure*}
\centering
\begin{picture}(500,200)(0,0)
\put(-5,0){
\begin{picture}(0,0) 
        \includegraphics[width=1.0\columnwidth]{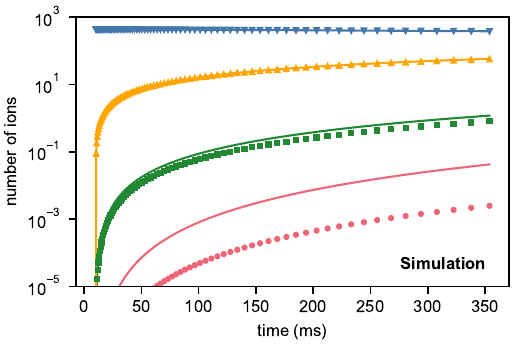}
\end{picture}}
\put(250,0){
\begin{picture}(0,0) 
        \includegraphics[width=1.0\columnwidth]{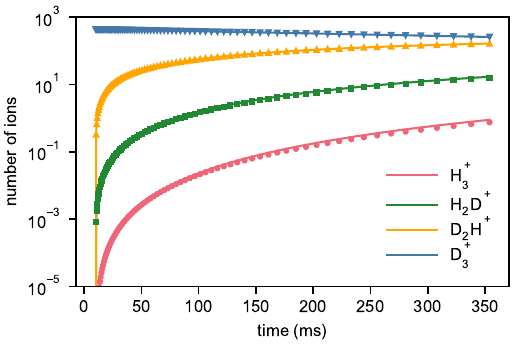}
\end{picture}}
\end{picture}  
    \caption{Number of ions as a function of reaction time in a simulation mimicking the \ce{D3+ + H2} experiment, 
    where the gas initially only consists 
    of \ce{D3+} and \ce{H2}. The results are shown for simulations run at 15\,K (left) and 50\,K (right). 
    The solid lines represent simulations using the averaged rate coefficients, and the markers represent a simulation where spin states 
    have been explicitly included, but summed over afterward.}
    \label{fig:opTests}
\end{figure*}

We tested the validity of the rate-coefficient averaging process by running test simulations where we either used the reduced 
reaction network or the original spin-state resolved network. In the latter case, we summed over the abundances of the spin forms 
of each species post-simulation so that they could be directly compared to the results obtained with the reduced network. 
Figure\,\ref{fig:opTests} shows the results of this comparison in physical conditions mimicking the \ce{D3+ + H2} experiment. 
The time evolution of the various species simulated with the full network, with explicit consideration of spin 
states\footnote{At the high-temperature limit, the \ce{D3+} ortho:meta:para ratio is 16:10:1.}, deviates from the solution 
derived using the reduced network at low temperatures. However, toward higher 
temperatures, the two approaches start to produce nearly identical results. The experiments presented here probed temperatures 
in excess of 20\,K, which justifies the use of the averaged rate coefficients. In any case, the set presented 
in Table\,\ref{tab:averagedRateCoeffs} only represents the initial guess, as we explain in more detail 
in Sect.\,\ref{ss:exampleSystem}.

We set the initial condition in each simulation by taking the recorded count of each ion at the earliest measurement 
time in the experimental data, and we used these values to define $t = 0$ in the simulation. In addition to the initial number 
of ions, the other parameters required for the chemical simulations are the number density of the reactants in the 
trap (in practice equal to the number density of the neutral reactant) and the temperature. Because pyRate works 
with number densities, we used the volume of the ion trap (0.770\,cm${^{-3}}$) to interconvert between ion counts 
and number densities when required.

We performed simulations of the \ce{H3+ + D2}, \ce{D2H+ + H2}, and \ce{D3+ + H2} reacting systems. 
In what follows, we discuss the \ce{H3+ + D2} system as a specific example and quantify the simulation process.

\subsection{Search for the best fit: \ce{H3+ + D2} system}\label{ss:exampleSystem}

The experiments yielded the number of ions as a function of storage time. Our simulations attempt to reproduce the time-dependent behavior of 
the experimental ion numbers. Because the experimentally derived rate coefficients indicate deviations from the predictions 
of \citet{Hugo2009}, we did not expect to find a match between the experiments and the simulations unless the initial rate coefficients
(Table\,\ref{tab:averagedRateCoeffs}) were modified. Therefore, we ran for each individual experiment 
an ensemble of $10^4$ simulations, 
where in each simulation the rate coefficient of every reaction was varied randomly by a factor of ten each way with respect to the 
initial guess. In this way, we searched for a set of rate coefficients that was able to provide a match to the experimental results. 
This approach is feasible because each simulation takes on the order of one second of CPU time to 
run\footnote{Most of this time was spent on initialization, data processing, and other related tasks. 
The solution of the ordinary differential equations takes only 
around one millisecond.}. For more time-consuming tasks, it would make more sense to use for example a Markov Chain Monte Carlo 
approach to search for the best-fitting solution.

The simulations were run from $t = 0 \, \rm s$ until approximately 10\,s even though the experimental measurements were 
typically taken over a duration of less than a second. To find the best-fitting simulation among the ensemble of $10^4$, 
we performed 
for each individual simulation a $\chi^2$ test on the simulated versus experimentally deduced number of ions for the four ions separately, 
for which we summed the four values to set the final $\chi^2$ value that represented the goodness of the fit. 
The best-fitting simulation 
is the one with the lowest $\chi^2$ value. We also derived error bars for the simulated rate coefficients by considering 
the simulations with a $\chi^2$ value within a factor of two from the minimum value to represent good fits. 
The value of the good-fit constraint is arbitrary and was set empirically based on examination of the simulation results. 
When we relaxed the condition and allowed more solutions within the good-fitting range, the distinction between sets of rate
coefficients were blurred, which provide an acceptable match to the experiments and those that only appear to do so due 
to the randomized nature of the simulations (see below for a more detailed discussion). In practice, with the constraint of a factor of 
two, typically only $5-10$ solutions for each set of 10,000 simulations were counted as good fits.

\begin{figure}
   \centering
   \includegraphics[width=\hsize]{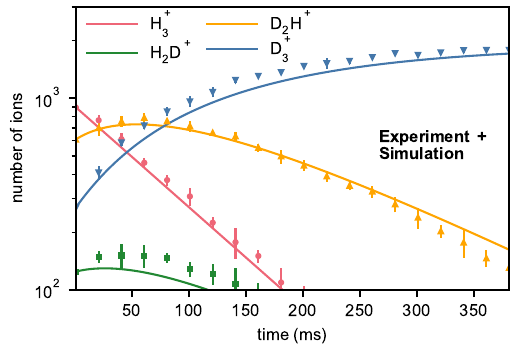}
      \caption{Number of the four ions as a function of reaction time in the \ce{H3+ + D2} experiment at a temperature 
      of 21.6\,K and a \ce{D2} number density of $\rm 9.83 \times 10^9 \, cm^{-3}$. The solid lines represent the best-fitting 
      simulation results, and the markers with error bars represent the experimental number of ions.}
         \label{fig:bestFitSimulation}
   \end{figure}

Figure~\ref{fig:bestFitSimulation} shows an example of the results, representing the best-fitting simulation of the \ce{H3+ + D2} system 
at a temperature of 21.6\,K. The agreement between the simulations and the experiment is very good. The maximum difference between 
the simulated and experimental number of ions is only some dozen percent. We also ran simulations for the other two \ce{D2} 
number density values considered in the experiments. The results were similar to those shown in Fig.\,\ref{fig:bestFitSimulation}, 
with a comparably good match between the experiments and the simulation. 
Additional information on the simulation of this system is presented in Appendix~\ref{appendix:Scaling}.

\subsection{Other systems and temperature dependence}

\begin{figure*}
   \centering
   \includegraphics[width=\hsize]{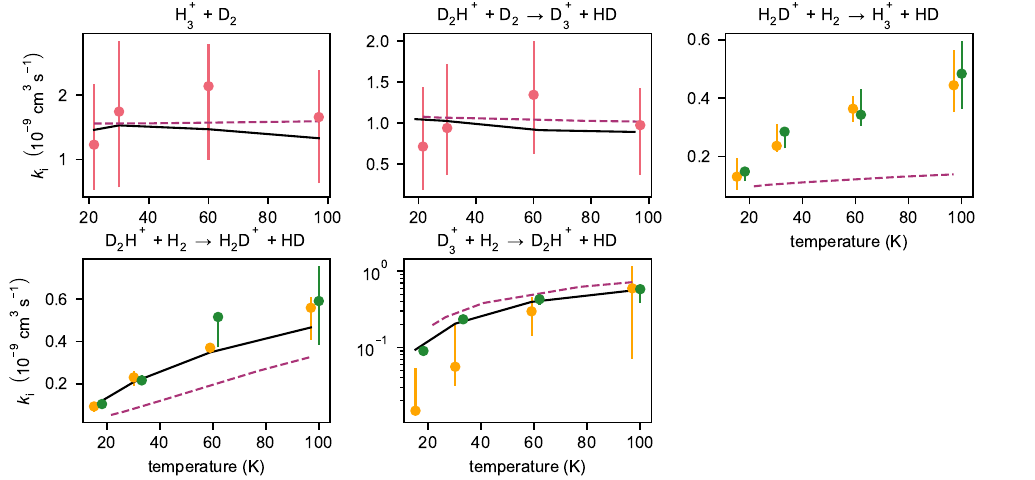}
      \caption{Rate coefficients of various reactions, as indicated above each panel, as a function of temperature. The filled 
      red, orange, and green circles indicate simulated rate coefficients based on the \ce{H3+ + D2}, \ce{D2H+ + H2}, 
      and \ce{D3+ + H2} systems, respectively. The error bars are defined as in Fig.\,\ref{fig:bestFitRateCoeffs}. The values 
      for the \ce{H3+ + D2} reaction are sums over the two product branches, that is, reactions 2 and 3 as given in
      Table\,\ref{tab:averagedRateCoeffs}. The rate coefficients of the \ce{H2D+ + H2 \to H3+ + HD}, \ce{D2H+ + H2 \to H2D+ + HD}, and \ce{D3+ + H2 \to D2H+ + HD} reactions are both constrained by the \ce{D2H+ + H2} and the \ce{D3+ + H2} 
      experimental systems and hence appear in two colors (\ce{D2H+ + H2} in orange; \ce{D3+ + H2} in green); the data have been offset 
      by 3\,K for better visibility of the error bars. The solid black lines indicate the experimentally deduced rate coefficients; 
      these data are missing for the 
      \ce{H2D+ + H2 \to H3+ + HD} reaction because they could not be determined experimentally. The dashed magenta lines 
      show the rate coefficient values as derived from the \citet{Hugo2009} data (see also Table\,\ref{tab:averagedRateCoeffs}).}
         \label{fig:collectedResultsFromAllSystems}
   \end{figure*}

Figure~\ref{fig:collectedResultsFromAllSystems} shows the temperature dependence of the simulated counterparts of all experimentally 
studied reactions in comparison with the actual experimental data. Simulations were performed for four temperature values ranging from 
approximately 20\,K to 100\,K. In general, the simulated values agree well with the experiments. 
For the \ce{H3+ + D2} reaction, which has two possible branches that could not be distinguished in the experiments, 
we compared the sum of the two rate coefficients to the experimental data. 
The \ce{D2H+ + H2} and \ce{D3+ + H2} reactions also have two possible product branches each, but in both 
cases, one branch has a very low rate coefficient (Table\,\ref{tab:averagedRateCoeffs}), and hence, we only show the major branch in 
Fig.\,\ref{fig:collectedResultsFromAllSystems}. Overall, the simulated and experimental rate coefficients 
deviate from the \citet{Hugo2009} values by a factor of a few at most, depending on the reaction and on the temperature.

The rate coefficients of the \ce{H2D+ + H2 \to H3+ + HD}, \ce{D2H+ + H2 \to H2D+ + HD}, and \ce{D3+ + H2 \to D2H+ + HD} reactions 
are both constrained by the \ce{D2H+ + H2} and the \ce{D3+ + H2} experiments. 
The figure shows that we obtained similar rate coefficients for the first two reactions regardless of the experiment 
on which the simulations were based, but for the last reaction, there is a clear difference between the simulation predictions. 
This is because the amount of $\rm D_3^+$ created in the \ce{D2H+ + H2} experiment is low compared to the other ions, and the simulations 
have difficulties to reproduce its number density. Thus, for the \ce{D3+ + H2 \to D2H+ + HD} reaction, 
we consider only the simulation predictions based on the \ce{D2H+ + H2} experimental system (green dots in the figure) to be reliable.

We were unable to measure the rate coefficient of the \ce{H2D+ + H2 \to H3+ + HD} reaction in the experiments.
This is the only reaction outside of the directly measured ones for which the simulations are able to give predictions with reasonably 
small error bars. The simulations indicate that the rate coefficient of this reaction was underestimated by \citet{Hugo2009} by a factor 
of approximately four at 100\,K, but, interestingly, the simulations also predict a pronounced temperature dependence, and the predicted rate 
coefficient value is very close to that of \citet{Hugo2009} at 20\,K. The astrochemical implications of this finding 
are discussed in Sect.\,\ref{ss:implications}, but we stress here that the simulation prediction should be corroborated by experiments 
with a different setup or by theoretical calculations.

The data shown in Fig.\,\ref{fig:collectedResultsFromAllSystems} correspond in each case to the lowest number density of the neutral 
reactant tested in the experiments. As demonstrated by Fig.\,\ref{fig:bestFitRateCoeffs}, some variation in the results can be expected 
when the density of the neutral reactant is varied. We verified via repeated simulations, however, that the randomness inherent 
to the simulation process (one obtains somewhat different values for the derived rate coefficients every time a new set of 
simulations is run) leads to variations in the results on the same order of magnitude as the changes in the neutral reactant density. 
Therefore, although the best-fit rate coefficient values change from one set of simulations to the next, the trends evident 
in Fig.\,\ref{fig:bestFitRateCoeffs} are not affected, and our conclusions remain the same regardless of the density of the neutral reactant.

\subsection{Conclusions from the simulations}

It is clear that for a given experimental system, the simulations do not allow us to constrain the rate coefficients of reactions 
that are not as important in that experiment, that is, we cannot constrain reactions whose rate coefficients are 
too low for them to contribute to the evolution of the experiment (or if the ion abundance is too low, as is the case for reactions 
7 and 8 in the \ce{H3+ + D2} experiment; see also Appendix~\ref{appendix:Scaling}). 
In the exceptional case of the \ce{H2D+ + H2 \to H3+ + HD} reaction, 
which could not be determined experimentally, we were able to establish limits on the reaction rate coefficient 
using the simulations, however. As it stands, our simulations serve to corroborate the experimental findings, as demonstrated by the good correspondence of 
the simulated rate coefficients to the experimental ones, but the error bars are generally too large to make reliable predictions 
on unmeasured reactions.

\section{Astrochemical implications}\label{ss:implications}

For the reactions involving \ce{D2} as the neutral reactant, the experiments predict very similar (within a few dozen percent) rate coefficients with 
respect to the values reported by \citet{Hugo2009}. 
However, larger differences of up to a factor of a few are found for reactions involving \ce{H2}, 
which are backward processes that lead to a reduction in deuterium substitution of the reacting 
$\rm H_3^+$ isotopolog. There is no consistent trend in the difference between our 
predictions and those of \citet{Hugo2009}, however, in that we find a higher rate coefficient for the \ce{D2H+ + H2 \to H2D+ + HD} reaction, 
but a lower one for \ce{D3+ + H2 \to D2H+ + HD}. Nevertheless, variations in the rate coefficients of these reactions imply 
effects on the relative populations of the deuterated forms of $\rm H_3^+$, which could have observational implications. 
The most striking feature of our simulations is the prediction for the rate coefficient of the \ce{H2D+ + H2 \to H3+ + HD} reaction, 
which we find to be higher by a factor of several at elevated temperatures in comparison to the value reported by \citet{Hugo2009} 
(the difference is about a factor of four at 100\,K), but at low temperatures, their data and ours agree very well. 
$\rm H_2D^+$ is not likely to be very abundant at elevated temperatures, and hence, the effect of the rate coefficient 
differences may be limited. On the other hand, however, it is conceivable that in the lukewarm temperatures of protostellar 
envelopes, a higher-than-expected destruction rate of $\rm H_2D^+$ might affect the overall deuteration degree. 
The overall implication of our results is that existing chemical models may be underestimating the efficiency of 
the destruction of two main deuterium carriers, $\rm H_2D^+$ and $\rm D_2H^+$. At the same time, it needs to be 
emphasized that we are unable with the current experimental setup to predict the rate coefficients of the forward 
processes (\eg, \ce{H3+ + HD \to H2D+ + H2}), and hence, it is difficult to draw any definite conclusions. It is clear, however, 
that further studies of these fundamental reactions are required for a complete understanding of deuterium chemistry in the interstellar medium.

The new values for the rate coefficients that the experiments provide are readily applicable in astrochemical models that do not 
distinguish spin states, with the exception of the \ce{H3+ + D2} reaction, for which the experiments cannot predict the branching 
ratio of the two possible product channels\footnote{The error bars in the simulations are also too large to draw definite conclusions on the branching ratio.}. 
For models that do include an explicit treatment of spin-state chemistry, the situation 
is more challenging as the spin states need to be introduced. Procedures for this are available in the literature 
\citep{Sipila15b,Hily-Blant18}, but it needs to be noted that the \ce{H3+ + H2} system is special in the sense that small 
energy differences between the reactants have a large impact on the relative populations, and the automated procedures 
presented in the literature may lead to false predictions for this system. We leave an in-depth investigation of this problem to future work.

Our experimental prediction for the rate coefficient of the \ce{H2+ + H2} reaction agrees very well with the data already available 
in KIDA, where no temperature dependence for this reaction is reported. The experimentally deduced kinetic isotope effect, with 
the rate coefficient decreasing with the amount of deuterium substitution (see Fig.~\ref{fig_exp}), or the temperature dependence, 
has not been taken into account to our knowledge in earlier gas-grain chemical models that dealt with deuterium chemistry. However, 
owing to the low abundance of \ce{D2}, as already pointed out above, these reactions are not likely to be very important for 
the overall evolution of deuterium chemistry in the ISM, and hence, variations on the order of a few dozen percent are 
not expected to change the conclusions drawn from chemical models significantly.

In stark contrast to \ce{D2}, helium is very abundant. Protonated helium (\ce{HeH+}) can be produced in the reaction 
between \ce{He+} and HCO \citep{Woon09}, which is currently the only \ce{HeH+} production path included in KIDA. 
Production of \ce{HeH+} via this pathway is inhibited in the ISM due to the higher ionization potential of helium 
compared to hydrogen, however, which means that \ce{He+} is exclusively produced by cosmic rays impacting on atomic helium. 
As a test, we ran a chemical simulation to estimate the \ce{HeH+} abundance in molecular clouds, assuming canonical 
values $n({\ce{H2}}) = 10^4\, \rm cm^{-3}$ and $T = 10\,\rm K$ for the number density and kinetic temperature, respectively. 
We find that the \ce{HeH+} abundance (with respect to \ce{H2}) is very low, on the order of $10^{-19} - 10^{-18}$. 
In the UMIST database \citep{Millar2024}, there is another \ce{HeH+} production pathway via \ce{H2+ + He} \citep{Theard74}, 
which is much more efficient in producing \ce{HeH+} compared to direct helium ionization followed by proton transfer. 
However, even with this reaction, we predict a \ce{HeH+} abundance of only a few times $10^{-14}$. Our tests were 
limited, but we find it likely that the potential of the \ce{HeD+ + H2} reaction as a means of distributing deuterium in 
the ISM remains low, despite the relatively high rate coefficient of this reaction (Fig.\,\ref{fig_exp_He_only}).

\section{Conclusions}

The rate coefficients of some of the deuterated variations of the \ce{H2+ + H2} and \ce{H3+ + H2} systems were 
experimentally determined for temperatures between 10 and 250~K. The processes involving diatomic ions were 
observed to follow the Langevin model and do not present a temperature dependence. The rates of the isotopic exchange processes
of the triatomic ions, on the other hand, were found to depend on the temperature regardless of whether they are exothermic or endothermic. 
Particularly for the endothermic hydrogenation processes, the internal energy of \ce{$o$-H2}, which is close to the 
endothermicity of the reactions, likely softens the drop in the measured rate coefficient with decreasing temperature. Nevertheless, a significant 
variation in the rate coefficient was obtained between 20 and 250~K, whether it is a decrease of $\sim 30\%$ for the exothermic processes
we studied or an increase of an order of magnitude for the endothermic processes.

The chemical simulations we used to fit the ion trap measurements provide rate coefficients that agree with the experimental 
coefficients, but struggle to give additional insight into most of the unmeasured rates due to the large error bars
obtained from the fit. 
Nevertheless, the simulations were able to constrain the rate coefficient for the reaction of \ce{H2D+ + H2}, 
and we obtained a stronger temperature dependence than the commonly used values of \cite{Hugo2009}.
This is in line with our other measurements for the hydrogenation of \ce{D2H+} and \ce{D3+}. This result highlights the need for a
direct measurement of this rate coefficient, which is lacking to date.

The rate coefficients obtained in this work can be readily included in astrochemical models that do not consider
state-to-state reactions, and they can be used over a wide range of temperatures.
Overall, our results show an enhanced efficiency of the hydrogenation of the partially deuterated ions
\ce{H2D+} and \ce{D2H+} compared to the values commonly used in astrochemical models. As a consequence, models may be overestimating the abundances of these ions and the relative importance of the deuteration processes
initiated by these species.

\begin{acknowledgements}
This work was supported by the Max Planck Society.
The authors gratefully acknowledge the work of the electrical and mechanical workshops and engineering 
departments of the Max Planck Institute for Extraterrestrial Physics.
We thank the reviewer for their constructive feedback.
\end{acknowledgements}

\subsection*{Data availability statement}
The data that support the findings of this study are openly available at {\tiny \url{https://doi.org/10.5281/zenodo.10777585}}.

\bibliographystyle{aa_url}
\bibliography{lit.bib}

\begin{thebibliography}{54}
\expandafter\ifx\csname natexlab\endcsname\relax\def\natexlab#1{#1}\fi

\bibitem[{Adams \& Smith(1981)}]{Adams1981}
Adams, N.~G. \& Smith, D. 1981,
  \href{http://dx.doi.org/10.1086/159162}{\color{magenta}\apj}, 248, 373

\bibitem[{Aguado {et~al.}(2010)Aguado, Barrag{\'a}n, Prosmiti,
  {Delgado-Barrio}, Villarreal, \& Roncero}]{Aguado2010}
Aguado, A., Barrag{\'a}n, P., Prosmiti, R., {et~al.} 2010,
  \href{http://dx.doi.org/10.1063/1.3454658}{\color{magenta}J. Chem. Phys.},
  133, 024306

\bibitem[{Allmendinger {et~al.}(2016{\natexlab{a}})Allmendinger, Deiglmayr,
  Höveler, Schullian, \& Merkt}]{Allmendinger2016b}
Allmendinger, P., Deiglmayr, J., Höveler, K., Schullian, O., \& Merkt, F.
  2016{\natexlab{a}},
  \href{http://dx.doi.org/10.1063/1.4972130}{\color{magenta}J. Chem. Phys.},
  145, 244316

\bibitem[{Allmendinger {et~al.}(2016{\natexlab{b}})Allmendinger, Deiglmayr,
  Schullian, Höveler, Agner, Schmutz, \& Merkt}]{Allmendinger2016a}
Allmendinger, P., Deiglmayr, J., Schullian, O., {et~al.} 2016{\natexlab{b}},
  \href{http://dx.doi.org/10.1002/cphc.201600828}{\color{magenta}ChemPhysChem},
  17, 3596

\bibitem[{Baer \& Ng(1990)}]{Baer1990}
Baer, M. \& Ng, C.~Y. 1990,
  \href{http://dx.doi.org/10.1063/1.459359}{\color{magenta}J. Chem. Phys.}, 93,
  7787

\bibitem[{{Barlow}(2004)}]{Barlow04}
{Barlow}, R. 2004,
  \href{https://ui.adsabs.harvard.edu/abs/2004physics...6120B}{\href{http://dx.doi.org/10.48550/arXiv.physics/0406120}{\color{magenta}arXiv
  e-prints}, physics/0406120}

\bibitem[{Bayet {et~al.}(2010)Bayet, Awad, \& Viti}]{Bayet2010}
Bayet, E., Awad, Z., \& Viti, S. 2010,
  \href{http://dx.doi.org/10.1088/0004-637X/725/1/214}{\color{magenta}ApJ},
  725, 214

\bibitem[{Caselli {et~al.}(2019)Caselli, Sipil{\"a}, \& Harju}]{Caselli2019}
Caselli, P., Sipil{\"a}, O., \& Harju, J. 2019,
  \href{http://dx.doi.org/10.1098/rsta.2018.0401}{\color{magenta}Phil. Trans.
  R. Soc. London, Ser. A}, 377, 20180401

\bibitem[{Caselli {et~al.}(2003)Caselli, van~der Tak, Ceccarelli, \&
  Bacmann}]{Caselli2003}
Caselli, P., van~der Tak, F. F.~S., Ceccarelli, C., \& Bacmann, A. 2003,
  \href{http://dx.doi.org/10.1051/0004-6361:20030526}{\color{magenta}A\&A},
  403, L37

\bibitem[{Dashevskaya {et~al.}(2016)Dashevskaya, Litvin, Nikitin, \&
  Troe}]{Dashevskaya2016}
Dashevskaya, E.~I., Litvin, I., Nikitin, E.~E., \& Troe, J. 2016,
  \href{http://dx.doi.org/10.1063/1.4972129}{\color{magenta}J. Chem. Phys.},
  145, 244315

\bibitem[{Douglass {et~al.}(1977)Douglass, McClure, \& Gentry}]{Douglass1977}
Douglass, C.~H., McClure, D.~J., \& Gentry, W.~R. 1977,
  \href{http://dx.doi.org/10.1063/1.434675}{\color{magenta}J. Chem. Phys.}, 67,
  4931

\bibitem[{Drozdovskaya {et~al.}(2021)Drozdovskaya, Schroeder~I, Rubin, Altwegg,
  {van~Dishoeck}, Kulterer, De~Keyser, Fuselier, \& Combi}]{Drozdovskaya2021}
Drozdovskaya, M.~N., Schroeder~I, I. R. H.~G., Rubin, M., {et~al.} 2021,
  \href{http://dx.doi.org/10.1093/mnras/staa3387}{\color{magenta}\mnras}, 500,
  4901

\bibitem[{Eaker \& Schatz(1985)}]{Eaker1985}
Eaker, C.~W. \& Schatz, G.~C. 1985,
  \href{http://dx.doi.org/10.1021/j100258a036}{\color{magenta}J. Phys. Chem.},
  89, 2612

\bibitem[{Gerlich(1992)}]{Gerlich1992}
Gerlich, D. 1992,
  \href{http://dx.doi.org/10.1002/9780470141397.ch1}{\color{magenta}Adv. Chem.
  Phys.}, LXXXII, 1

\bibitem[{Gerlich {et~al.}(2002)Gerlich, Herbst, \& Roueff}]{Gerlich2002a}
Gerlich, D., Herbst, E., \& Roueff, E. 2002,
  \href{http://dx.doi.org/10.1016/S0032-0633(02)00094-6}{\color{magenta}P\&SS},
  50, 1275

\bibitem[{Gerlich {et~al.}(2013)Gerlich, Pla\v{s}il, Zymak, Hejduk, Jusko,
  Mulin, \& Glos\'ik}]{Gerlich2013}
Gerlich, D., Pla\v{s}il, R., Zymak, I., {et~al.} 2013,
  \href{http://dx.doi.org/10.1021/jp400917v}{\color{magenta}J. Phys. Chem. A},
  117, 10068

\bibitem[{Gerlich \& Schlemmer(2002)}]{Gerlich2002}
Gerlich, D. \& Schlemmer, S. 2002,
  \href{http://dx.doi.org/10.1016/S0032-0633(02)00095-8}{\color{magenta}P\&SS},
  50, 1287

\bibitem[{Gerlich {et~al.}(2006)Gerlich, Windisch, Hlavenka, Pla{\v s}il, \&
  Glosik}]{Gerlich2006}
Gerlich, D., Windisch, F., Hlavenka, P., Pla{\v s}il, R., \& Glosik, J. 2006,
  \href{http://dx.doi.org/10.1098/rsta.2006.1865}{\color{magenta}Phil. Trans.
  R. Soc. London, Ser. A}, 364, 3007

\bibitem[{Giles {et~al.}(1992)Giles, Adams, \& Smith}]{Giles1992}
Giles, K., Adams, N.~G., \& Smith, D. 1992,
  \href{http://dx.doi.org/10.1021/j100198a030}{\color{magenta}J. Phys. Chem.},
  96, 7645

\bibitem[{Glenewinkel-Meyer \& Gerlich(1997)}]{Glenewinkel1997}
Glenewinkel-Meyer, T. \& Gerlich, D. 1997,
  \href{http://dx.doi.org/10.1002/ijch.199700039}{\color{magenta}Isr. J.
  Chem.}, 37, 343

\bibitem[{Glos{\'i}k(1994)}]{Glosik1994}
Glos{\'i}k, J. 1994,
  \href{http://dx.doi.org/10.1016/0168-1176(94)90004-3}{\color{magenta}Int. J.
  Mass Spectrom. Ion Processes}, 139, 15

\bibitem[{{G{\'o}mez-Carrasco} {et~al.}(2012){G{\'o}mez-Carrasco},
  {Gonz{\'a}lez-S{\'a}nchez}, Aguado, {Sanz-Sanz}, Zanchet, \&
  Roncero}]{Gomez-Carrasco2012}
{G{\'o}mez-Carrasco}, S., {Gonz{\'a}lez-S{\'a}nchez}, L., Aguado, A., {et~al.}
  2012, \href{http://dx.doi.org/10.1063/1.4747548}{\color{magenta}J. Chem.
  Phys.}, 137, 094303

\bibitem[{{Hily-Blant} {et~al.}(2018){Hily-Blant}, {Faure}, {Rist}, {Pineau des
  For{\^e}ts}, \& {Flower}}]{Hily-Blant18}
{Hily-Blant}, P., {Faure}, A., {Rist}, C., {Pineau des For{\^e}ts}, G., \&
  {Flower}, D.~R. 2018,
  \href{http://dx.doi.org/10.1093/mnras/sty881}{\color{magenta}\mnras},
  \href{http://adsabs.harvard.edu/abs/2018MNRAS.477.4454H}{477, 4454}

\bibitem[{H{\"o}veler {et~al.}(2021{\natexlab{a}})H{\"o}veler, Deiglmayr,
  Agner, Schmutz, \& Merkt}]{Hoveler2021a}
H{\"o}veler, K., Deiglmayr, J., Agner, J.~A., Schmutz, H., \& Merkt, F.
  2021{\natexlab{a}},
  \href{http://dx.doi.org/10.1039/D0CP06107G}{\color{magenta}Phys. Chem. Chem.
  Phys.}, 23, 2676

\bibitem[{H{\"o}veler {et~al.}(2021{\natexlab{b}})H{\"o}veler, Deiglmayr, \&
  Merkt}]{Hoveler2021b}
H{\"o}veler, K., Deiglmayr, J., \& Merkt, F. 2021{\natexlab{b}},
  \href{http://dx.doi.org/10.1080/00268976.2021.1954708}{\color{magenta}Mol.
  Phys.}, 119, e1954708

\bibitem[{Hugo {et~al.}(2009)Hugo, Asvany, \& Schlemmer}]{Hugo2009}
Hugo, E., Asvany, O., \& Schlemmer, S. 2009,
  \href{http://dx.doi.org/10.1063/1.3089422}{\color{magenta}J. Chem. Phys.},
  130, 164302

\bibitem[{Jiménez-Redondo {et~al.}(2011)Jiménez-Redondo, Carrasco, Herrero,
  \& Tanarro}]{Redondo2011}
Jiménez-Redondo, M., Carrasco, E., Herrero, V.~J., \& Tanarro, I. 2011,
  \href{http://dx.doi.org/10.1039/C1CP20426B}{\color{magenta}Phys. Chem. Chem.
  Phys.}, 13, 9655

\bibitem[{Jusko {et~al.}(2024)Jusko, Jim\'{e}nez-Redondo, \&
  Caselli}]{Jusko2023}
Jusko, P., Jim\'{e}nez-Redondo, M., \& Caselli, P. 2024,
  \href{http://dx.doi.org/10.1080/00268976.2023.2217744}{\color{magenta}Mol.
  Phys.}, 122, e2217744

\bibitem[{Krenos {et~al.}(1976)Krenos, Lehmann, Tully, Hierl, \&
  Smith}]{Krenos1976}
Krenos, J.~R., Lehmann, K.~K., Tully, J.~C., Hierl, P.~M., \& Smith, G.~P.
  1976,
  \href{http://dx.doi.org/10.1016/0301-0104(76)89028-3}{\color{magenta}Chem.
  Phys.}, 16, 109

\bibitem[{McCall \& Oka(2000)}]{McCall2000}
McCall, B.~J. \& Oka, T. 2000,
  \href{http://dx.doi.org/10.1126/science.287.5460.1941}{\color{magenta}Sci},
  287, 1941

\bibitem[{Merkt {et~al.}(2022)Merkt, Höveler, \& Deiglmayr}]{Merkt2022}
Merkt, F., Höveler, K., \& Deiglmayr, J. 2022,
  \href{http://dx.doi.org/10.1021/acs.jpclett.1c03374}{\color{magenta}J. Phys.
  Chem. Lett.}, 13, 864

\bibitem[{Milenko {et~al.}(1972)Milenko, Karnatsevich, \& Kogan}]{Milenko1972}
Milenko, Y., Karnatsevich, L., \& Kogan, V. 1972,
  \href{http://dx.doi.org/10.1016/0031-8914(72)90223-6}{\color{magenta}Physica},
  60, 90

\bibitem[{Millar(2002)}]{Millar2002}
Millar, T.~J. 2002,
  \href{http://dx.doi.org/10.1016/S0032-0633(02)00082-X}{\color{magenta}P\&SS},
  50, 1189

\bibitem[{Millar {et~al.}(1989)Millar, Bennett, \& Herbst}]{Millar1989}
Millar, T.~J., Bennett, A., \& Herbst, E. 1989,
  \href{http://dx.doi.org/10.1086/167444}{\color{magenta}\apj}, 340, 906

\bibitem[{Millar {et~al.}(2024)Millar, Walsh, de~Sande, \&
  Markwick}]{Millar2024}
Millar, T.~J., Walsh, C., de~Sande, M.~V., \& Markwick, A.~J. 2024,
  \href{http://dx.doi.org/10.1051/0004-6361/202346908}{\color{magenta}A\&A},
  682, A109

\bibitem[{Oka(2013)}]{Oka2013}
Oka, T. 2013, \href{http://dx.doi.org/10.1021/cr400266w}{\color{magenta}Chem.
  Rev.}, 113, 8738

\bibitem[{Paul {et~al.}(1996)Paul, Schlemmer, L{\"u}cke, \& Gerlich}]{Paul1996}
Paul, W., Schlemmer, S., L{\"u}cke, B., \& Gerlich, D. 1996,
  \href{http://dx.doi.org/10.1016/0301-0104(96)00160-7}{\color{magenta}Chem.
  Phys.}, 209, 265

\bibitem[{Pollard {et~al.}(1991)Pollard, Johnson, Lichtin, \&
  Cohen}]{Pollard1991}
Pollard, J.~E., Johnson, L.~K., Lichtin, D.~A., \& Cohen, R.~B. 1991,
  \href{http://dx.doi.org/10.1063/1.461704}{\color{magenta}J. Chem. Phys.}, 95,
  4877

\bibitem[{{Ramanlal, J.} {et~al.}(2003){Ramanlal, J.}, {Polyansky, O. L.}, \&
  {Tennyson, J.}}]{Ramanlal2003}
{Ramanlal, J.}, {Polyansky, O. L.}, \& {Tennyson, J.} 2003,
  \href{http://dx.doi.org/10.1051/0004-6361:20030774}{\color{magenta}A\&A},
  406, 383

\bibitem[{{Sanz-Sanz} {et~al.}(2015){Sanz-Sanz}, Aguado, Roncero, \&
  Naumkin}]{Sanz-Sanz2015}
{Sanz-Sanz}, C., Aguado, A., Roncero, O., \& Naumkin, F. 2015,
  \href{http://dx.doi.org/10.1063/1.4937138}{\color{magenta}J. Chem. Phys.},
  143, 234303

\bibitem[{Savić {et~al.}(2020)Savić, Schlemmer, \& Gerlich}]{Savic2020}
Savić, I., Schlemmer, S., \& Gerlich, D. 2020,
  \href{http://dx.doi.org/10.1002/cphc.202000258}{\color{magenta}ChemPhysChem},
  21, 1429

\bibitem[{Shimonishi {et~al.}(2021)Shimonishi, Izumi, Furuya, \&
  Yasui}]{Shimonishi2021}
Shimonishi, T., Izumi, N., Furuya, K., \& Yasui, C. 2021,
  \href{http://dx.doi.org/10.3847/1538-4357/ac289b}{\color{magenta}ApJ}, 922,
  206

\bibitem[{{Sipil{\"a}} {et~al.}(2015{\natexlab{a}}){Sipil{\"a}}, {Caselli}, \&
  {Harju}}]{Sipila15a}
{Sipil{\"a}}, O., {Caselli}, P., \& {Harju}, J. 2015{\natexlab{a}},
  \href{http://dx.doi.org/10.1051/0004-6361/201424364}{\color{magenta}\aap},
  \href{http://adsabs.harvard.edu/abs/2015A%26A...578A..55S}{578, A55}

\bibitem[{{Sipil{\"a}} {et~al.}(2017){Sipil{\"a}}, {Harju}, \&
  {Caselli}}]{Sipila17b}
{Sipil{\"a}}, O., {Harju}, J., \& {Caselli}, P. 2017,
  \href{http://dx.doi.org/10.1051/0004-6361/201731039}{\color{magenta}\aap},
  \href{http://adsabs.harvard.edu/abs/2017A%26A...607A..26S}{607, A26}

\bibitem[{{Sipil{\"a}} {et~al.}(2015{\natexlab{b}}){Sipil{\"a}}, {Harju},
  {Caselli}, \& {Schlemmer}}]{Sipila15b}
{Sipil{\"a}}, O., {Harju}, J., {Caselli}, P., \& {Schlemmer}, S.
  2015{\natexlab{b}},
  \href{http://dx.doi.org/10.1051/0004-6361/201526468}{\color{magenta}\aap},
  \href{http://adsabs.harvard.edu/abs/2015A%26A...581A.122S}{581, A122}

\bibitem[{Suleimanov {et~al.}(2018)Suleimanov, Aguado, {G{\'o}mez-Carrasco}, \&
  Roncero}]{Suleimanov2018}
Suleimanov, Y.~V., Aguado, A., {G{\'o}mez-Carrasco}, S., \& Roncero, O. 2018,
  \href{http://dx.doi.org/10.1021/acs.jpclett.8b00783}{\color{magenta}J. Phys.
  Chem. Lett.}, 9, 2133

\bibitem[{Tennyson(1995)}]{Tennyson1995}
Tennyson, J. 1995,
  \href{http://dx.doi.org/10.1088/0034-4885/58/4/002}{\color{magenta}Rep. Prog.
  Phys.}, 58, 421

\bibitem[{{Theard} \& {Huntress}(1974)}]{Theard74}
{Theard}, L.~P. \& {Huntress}, W.~T. 1974,
  \href{http://dx.doi.org/10.1063/1.1681453}{\color{magenta}\jcp},
  \href{https://ui.adsabs.harvard.edu/abs/1974JChPh..60.2840T}{60, 2840}

\bibitem[{Verhaegh {et~al.}(2021)Verhaegh, Lipschultz, Harrison, Duval, Fil,
  Wensing, Bowman, Gahle, Kukushkin, Moulton, Perek, Pshenov, Federici,
  F{\'e}vrier, Myatra, Smolders, Theiler, Team, \& Team}]{Verhaegh2021}
Verhaegh, K., Lipschultz, B., Harrison, J.~R., {et~al.} 2021,
  \href{http://dx.doi.org/10.1088/1741-4326/ac1dc5}{\color{magenta}Nucl.
  Fusion}, 61, 106014

\bibitem[{Wakelam {et~al.}(2012)Wakelam, Herbst, Loison, Smith, Chandrasekaran,
  Pavone, Adams, Bacchus-Montabonel, Bergeat, Béroff, Bierbaum, Chabot,
  Dalgarno, van Dishoeck, Faure, Geppert, Gerlich, Galli, Hébrard, Hersant,
  Hickson, Honvault, Klippenstein, Picard, Nyman, Pernot, Schlemmer, Selsis,
  Sims, Talbi, Tennyson, Troe, Wester, \& Wiesenfeld}]{Wakelam2012}
Wakelam, V., Herbst, E., Loison, J.-C., {et~al.} 2012,
  \href{http://dx.doi.org/10.1088/0067-0049/199/1/21}{\color{magenta}\apjs},
  199, 21

\bibitem[{Walmsley \& Flower(2004)}]{Walmsley2004}
Walmsley, C.~M. \& Flower, D.~R. 2004,
  \href{http://dx.doi.org/10.1051/0004-6361:20035718}{\color{magenta}A\&A},
  418, 1035

\bibitem[{Willacy \& Woods(2009)}]{Willacy2009}
Willacy, K. \& Woods, P.~M. 2009,
  \href{http://dx.doi.org/10.1088/0004-637X/703/1/479}{\color{magenta}ApJ},
  703, 479

\bibitem[{{Woon} \& {Herbst}(2009)}]{Woon09}
{Woon}, D.~E. \& {Herbst}, E. 2009,
  \href{http://dx.doi.org/10.1088/0067-0049/185/2/273}{\color{magenta}\apjs},
  \href{https://ui.adsabs.harvard.edu/abs/2009ApJS..185..273W}{185, 273}

\bibitem[{Xie {et~al.}(2005)Xie, Braams, \& Bowman}]{Xie2005}
Xie, Z., Braams, B.~J., \& Bowman, J.~M. 2005,
  \href{http://dx.doi.org/10.1063/1.1927529}{\color{magenta}J. Chem. Phys.},
  122, 224307

\end{thebibliography}

\begin{appendix}

\section{Reactions of \ce{HeH+}/\ce{HeD+} with \ce{H2}}\label{appendix:HeX}

\begin{figure}[h]
\includegraphics[]{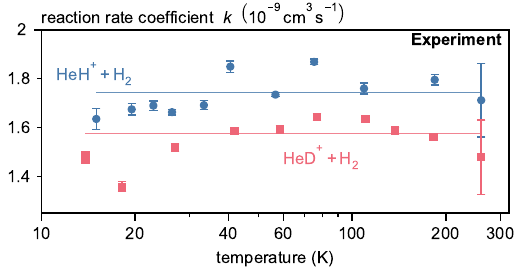}
\caption{Rate coefficients as a function of temperature for the \ce{HeH+} and \ce{HeD+} reactions with \ce{H2} studied in the 
  22 pole trap experiments. 
  Symbols correspond to the measurements, while lines represent the fit to the Arrhenius-Kooij formula with the parameters given 
  in Table\,\ref{tab:kida}.}
\label{fig_exp_He_only}
\end{figure}

As mentioned in the main text, \ce{HeH+} and \ce{HeD+} are 
co-trapped in the experiments involving diatomic ions. Due to overlapping masses 
(same mass-to-charge ratio $m/z$) only the determination of the rate coefficient for
\ce{HeH+ + H2} and \ce{HeD+ + H2} is possible. The corresponding rate coefficients
as a function of temperature are displayed in Fig.~\ref{fig_exp_He_only}.

\section{Scaling coefficients derived in the simulation of the $\rm H_3^+ + D_2$ system}\label{appendix:Scaling}

\begin{figure*}[!b]
   \centering
\includegraphics[width = 1.0\hsize]{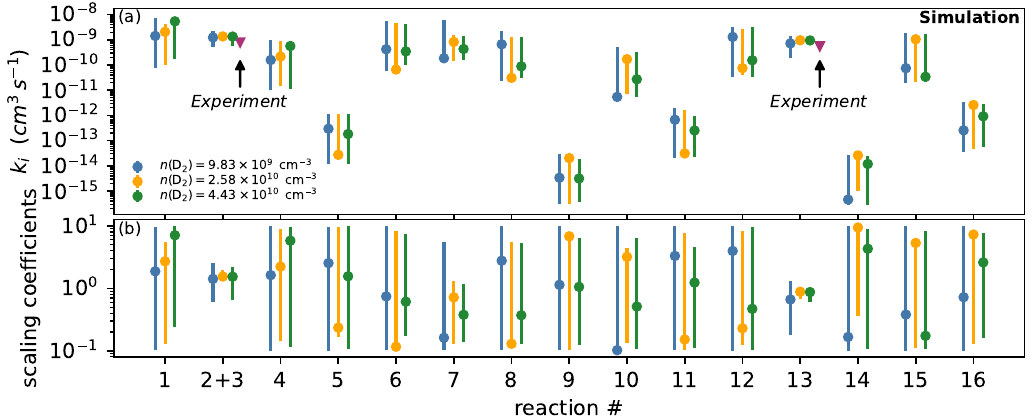}
      \caption{Data collected from the best-fitting simulations of the \ce{H3+ + D2} experiment at 21.6\,K. 
      The reaction number refers to the list in Table\,\ref{tab:averagedRateCoeffs}. Results are shown for three 
      sets of simulations corresponding to the three different variations of the experiment, where 
      the number density of the neutral reactant (\ce{D2}) 
      is varied as shown in the Legend. The results of the three simulation sets have been shifted from each other, 
      and are shown with unique colors, to better discriminate between them. The panels show the rate coefficients 
      of the 16 individual reactions (top) 
      and the scaling coefficients (bottom), which measure the factor of increase/decrease relative to the 
      fiducial rate coefficients that 
      was required to obtain the fit, for each reaction rate coefficient corresponding to the set of best-fitting simulations. 
      Purple triangles in the top panel mark the experimentally deduced rate coefficients. The data corresponding to 
      reactions~2~and~3 have been summed over so that a better correspondence to the experiments, which do not discriminate between
      the product branches, can be met. Error bars are not visible if they are smaller than the symbol, or, in the case of 
      the summed-up reactions 2 and 3, due to the fact that error bars are not defined in each direction which prevents 
      the summation method of \citet{Barlow04} being used in a well-defined manner.}
         \label{fig:bestFitRateCoeffs}
\end{figure*}

To complement the discussion in Sect.\,\ref{ss:exampleSystem}, we show in Fig.\,\ref{fig:bestFitRateCoeffs} a breakdown 
of the predicted rate coefficients of the good-fitting solutions, obtained via the random scaling process, for 
the $\rm H_3^+ + D_2$ system, for each of the 16 reactions individually. For this pair of reactants, only the 
reactions 2, 3, 7, 8, and 13 (Table~\ref{tab:averagedRateCoeffs}) are of practical consequence because low amounts 
of \ce{H2} and HD are formed during the experiment with respect to \ce{D2}. Furthermore, the \ce{H2D+} count is 
low (see Fig.\,\ref{fig:bestFitSimulation}), meaning that reactions 7 and 8 are not very important to the overall 
evolution of the system either. These facts are reflected in the scaling coefficients associated with the 
good-fitting simulations; only the rate coefficients of reactions 2, 3, and 13 are constrained in a manner that can 
be considered reliable -- all of the other rate coefficients present a large amount of scatter. The experiments 
do not distinguish between branching ratios when there are multiple product channels but instead yield the total 
rate coefficient, and hence we have added up the simulated rate coefficients of reactions 2 and 3 so that a direct 
comparison between the simulations and the experiment is possible. The summation was done using 
the ``variable Gaussian'' method \citet{Barlow04} allowing error bars to be obtained.

\end{appendix}

\end{document}